# Machine-learning enables Image reconstruction and classification in a "see-through" camera

ZHIMENG PAN,[1] BRIAN RODRIGUEZ,[2,3] AND RAJESH MENON[2,*]

[1]*School of Computing, University of Utah, 50 Central Campus Dr. Salt Lake City UT 84112, USA*
[2]*Department of Electrical and Computer Engineering, University of Utah, 50 Central Campus Dr. Salt Lake City UT 84112, USA*
[2]*Currently with Banjo, Inc. Park City, UT.*
*\*rmenon@eng.utah.edu*

**Abstract:** We demonstrate that image reconstruction can be achieved via a convolutional neural network for a "see-through" computational camera comprised of a transparent window and CMOS image sensor. Furthermore, we compared classification results using a classifier network for the raw sensor data vs the reconstructed images. The results suggest that similar classification accuracy is likely possible in both cases with appropriate network optimizations. All networks were trained and tested for the MNIST (6 classes), EMNIST and the Kanji49 datasets.

## 1. Introduction

Imaging is a form of information transfer from the object to the image planes. The traditional camera comprised of lenses and an image sensor enables an approximately one-to-one mapping between these planes. This approach is widely successful primarily because of the high signal-to-noise ratio (SNR) that may be achieved at each image pixel. However, there are alternative one-to-many mappings that can achieve information transfer albeit with constraints. Such an approach could be useful for spectral imaging with no absorption losses [1,2], imaging in the angular-spectral dimensions [2] and also for imaging in restricted environments, such as microscopy within the brain of a mouse [4-7]. More recently, we have also demonstrated imaging with only the bare image sensor [8]. In this paper, we are only concerned with incoherent imaging, since that is the most general and particularly useful modality for imaging. We note that in all cases, the point-spread function of our system is space variant. In all previous examples, the images are reconstructed for human consumption via regularization-based matrix inversion following an experimental calibration step. Recently, we have shown that similar results could be achieved with machine learning as well [9]. In this paper, we explore two new aspects of a recently demonstrated "see-through" or transparent camera [10]: first, we show that a trained neural network is able to perform image reconstruction from such a camera; second, we explore the difference between image reconstruction and image classification in such a camera, a problem that was deemed to be most interesting and least studied in a recent review on machine-learning-based imaging [11].

As was described before, our "see-through" camera is comprised of an image sensor placed at the edge of a transparent window [10]. A schematic and photograph of our experimental setup are shown in Fig. 1. The object was a conventional LCD display, the window was made of transparent plexiglass and the sensor was a color CMOS image sensor (MU300 from AmScope). The distance between the window and the LCD was approximately 250mm. The test images were displayed on the LCD and the corresponding sensor data was captured and stored. Ten frames were averaged for each stored data frame to reduce noise. A black box was used to cover the setup to minimize any ambient stray light.

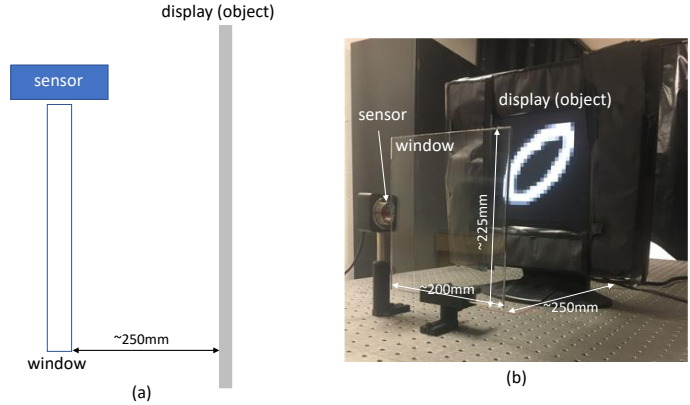

*Figure 1: (a) Schematic of our experimental setup. The object is an LCD display placed about 250mm away from a transparent plexiglass window, to the edge of which is placed a color CMOS image sensor (with no optics). (b) Photograph of our experimental setup. The letter "o" from the MNIST dataset is displayed on the LCD.*

## 2. Network Architecture and Training Methodologies

<u>Network for Image Reconstruction:</u> We built a convolutional neural network (CNN) to learn the inverse function that could reconstruct images from their corresponding sensor images. The overall network structure follows the classic "Unet" architecture [12], and modified with additional dense blocks from 'Res-Net' [13]. U net is a kind of 'encoder-decoder' architecture, where the input images first go through a series of stages of convolutional and pooling layers. Each stage will reduce the dimensions (height X width) of images by half, but doubles the number of channels. This phase is referred to as the encoding phase. After this phase, input images are encoded into a lower dimensional representation space. Then the encoded representations go through a decoder phase, which is very similar to the encoder phase. The difference is that each decoder stage will double the dimensions and halve the number of channels. The characteristic feature of U-net is the skip connection, which concatenates the corresponding encoder stage outputs to decoder stage inputs. By doing so the network could use as much of the original information as possible to reconstruct the images.

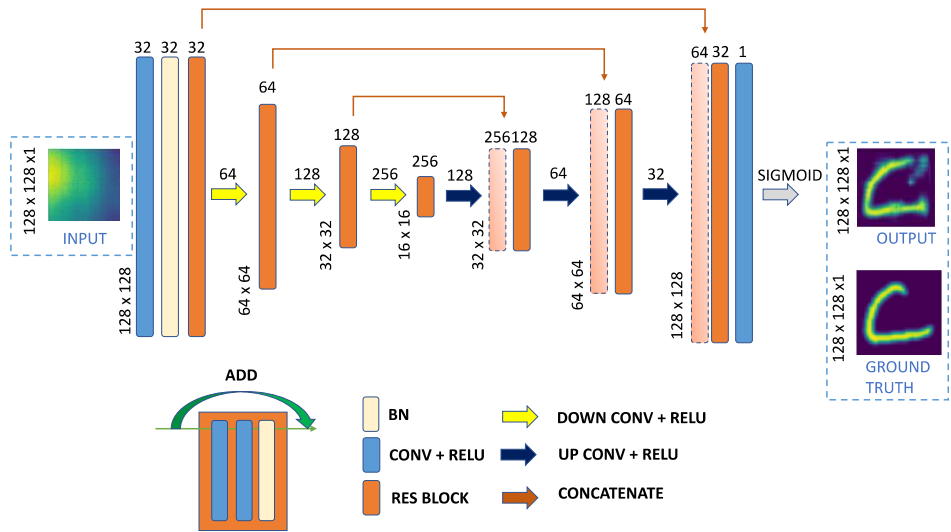

*Figure 2: CNN architecture for image reconstruction.*

Our dense block consists of 3 individual layers: 2 convolutional layers with RELU activation function followed by a batch-normalization layer. The advantage of the dense block is it could prevent the gradient from vanishing so that we could train very deep networks efficiently. Figure 2 shows the detailed architecture of the image-reconstruction CNN. Given the structure, the activation function of the last layer and the loss function are worth carefully considering. It has been well-known that the commonly used MSE, *i.e.*, mean-square error loss function does not work well in sparse-image reconstruction as it tends to produce blurred images [14]. Instead, we use the pixel-wise cross-entropy as the loss function (*L*), which could impose sparsity [15],

$$L = \frac{1}{N}\sum_i -g_i log(p_i) - (1 - g_i)log(1 - p_i),$$

where the summation is over every pixel *i*, and $g_i$ and $p_i$ represent the ground truth and predicted pixel intensity, respectively. In order to make the loss function valid, we need to restrict the range of the output layer to be within [0,1]. We thus choose sigmoid to be the activation function of the output layer. For all data sets, we split them into training set and testing set in the ratio of 9:1. We use the Adam optimizer [16] with initial learning rate 0.001, and train up to 50 epochs.

Network for Image Classification: Reconstructed results can be evaluated by difference functions like mean square error (MSE) or mean absolute error (MAE), or visual comparison. We will provide these metrics in the next section. In addition, we measure the quality of the reconstructed images by doing classification using the reconstructed images, and compare it to classification with original images and with raw sensor images (without reconstruction). Since our main goal of the classification is to test the reconstruction ability of the network in Fig. 2, rather than coming up with a state-of-the-art classification network, we decided to use an off-the-shelf classification network, SimpleNet [17]. Though simple, having the fewest parameters compared with other architectures, SimpleNet has proved to provide very competitive, sometimes offering better performance in classification tasks.

Since the dataset and sensor image resolution may vary, for efficiency and consistency, we cropped the input and ground truth images to ratio 1:1 and scaled them to 128 x 128 pixels without anti-aliasing. We also manually added Gaussian white noise of mean = 0, variance = 0.001 to ground truth images to improve the robustness of our network. For classification of reconstructed images, we first converted the reconstructed images into 32x32 pixels, which is the resolution of original images and then fed them into the classification network. For the raw-sensor-data-based classification, we preserve the image aspect ratio and resized the image to 125 X 170 pixels.

## 3. Results and Discussion

We trained and tested the network from Fig. 2 on 3 data sets: MNIST (6 classes) [18], EMNIST [19] and Kanji49 [20]. MNIST is the most widely used data set for visual task, containing gray scale images of 10 digits in various handwritten forms. But using as a proof of concept, we only use the first 6 classes (0 to 5) and randomly sub-sample 10% images in each class. EMNIST is an augmentation of MNIST, which additionally contains handwritten images of 26 English alphabet characters for both upper and lower cases. Note that some characters (for example, x,y,z) have very similar forms for both cases and are thus merged into one class (see [19] for details). Therefore, the total number of classes in EMNIST is 47, instead of 62. Kanji49 is similar to MNIST, but instead contains 49 Japanese Hiragana characters. It basically has the same number of classes as EMNIST, but the shapes of Hiragana are more complicated than those of the English characters. We include this dataset to further verify the reconstruction ability of our network. Table 1 lists the summary of data sets we used.

**Table 1. Details of datasets and summary of results**

| Name | # of images | # of classes | Training MAE | Testing MAE |
|---|---|---|---|---|
| MNIST | 6,000 | 6 | 0.0129 | 0.1014 |
| EMNIST | 120,000 | 47 | 0.0730 | 0.1213 |
| KANJI49 | 100,000 | 49 | 0.0994 | 0.1786 |

For each dataset, we trained our model as described earlier. Table 1 contains reconstruction performance for both training and testing sets. Figures 3-5 show the reconstruction results for the MNIST, EMNIST and Kanji49 datasets, respectively. Both training and testing results are included. Note that the input, ground truth and output are all grayscale images. Reconstruction results of MNIST is the best, as expected, since it contains fewest number of classes. Kanji49's result is slight worse than EMNIST because the characters in it have more variants and are more difficult for the network to find a suitable inverse function. For all 3 datasets, the testing results are worse than training results as expected. This is a typical phenomenon in deep learning based algorithm especially when the volume of training samples is not big enough. In this case the network tends to over-fit the training set, so as to decrease the training loss. It's worth noting that we could always over-fit the training set with large enough network and bigger training batch. We could, to some extent, control the trade-off between small generalization gap (which means relatively high testing accuracy) and high training accuracy.

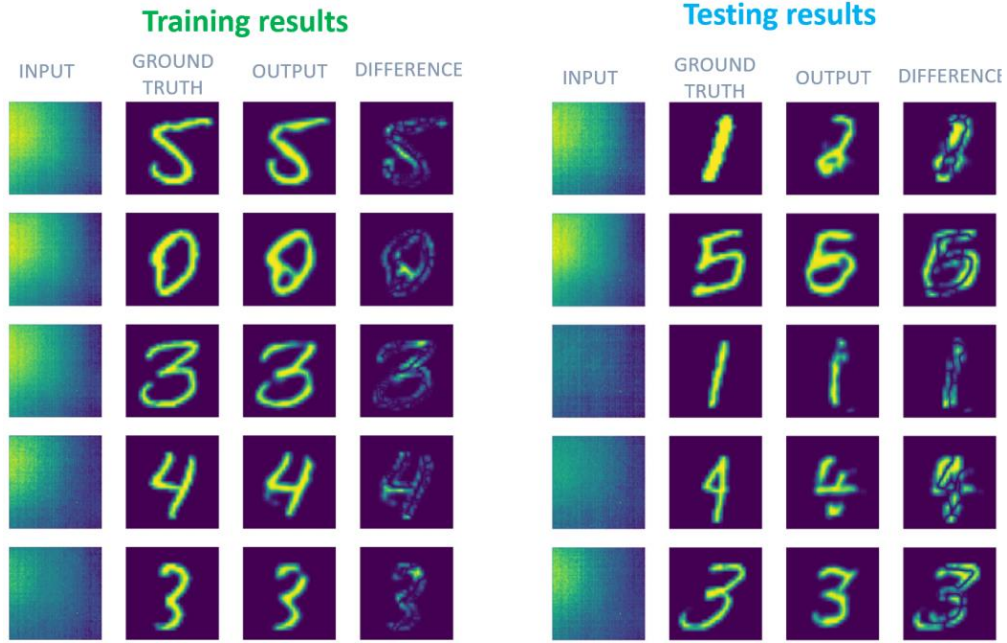

*Figure 3:* *Reconstruction results for MNIST data. Left shows example images from the training set and Right shows example images from the testing data set.*

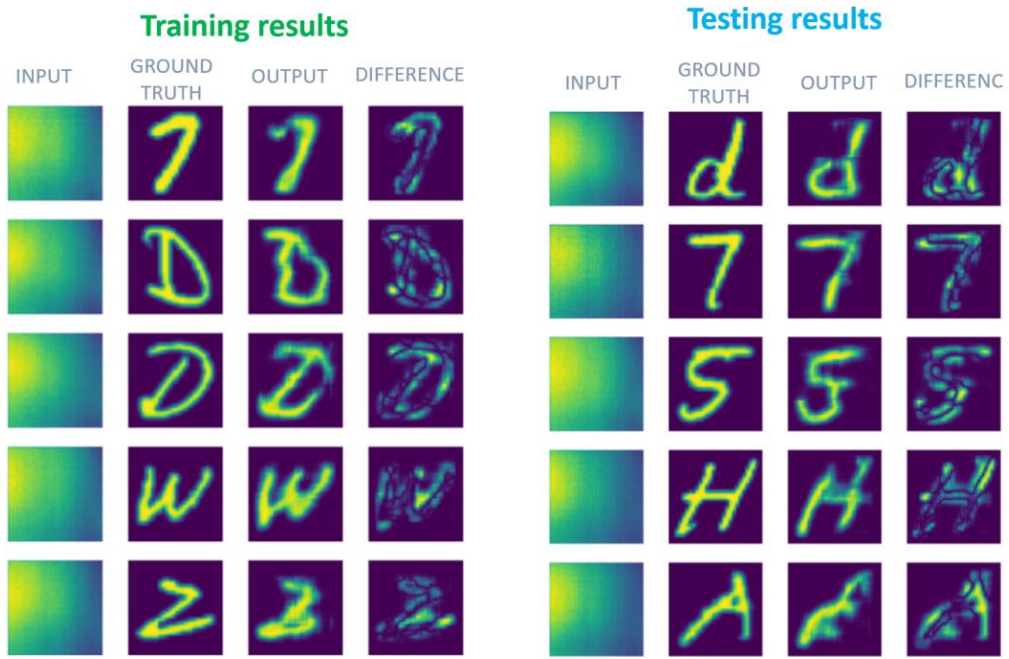

*Figure 4: Reconstruction results for EMNIST data. Left shows example images from the training set and Right shows example images from the testing data set.*

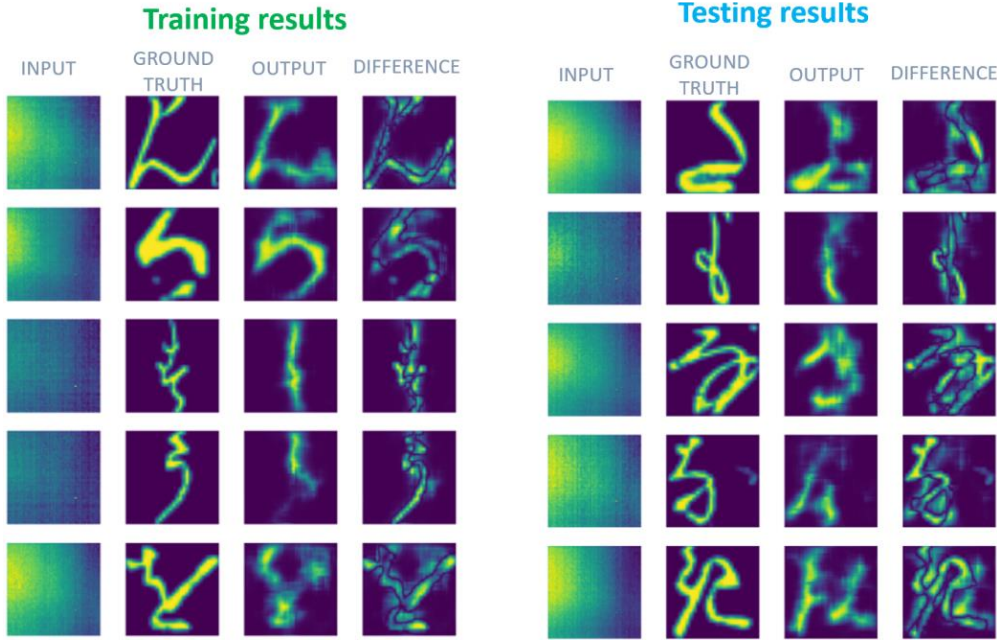

*Figure 5: Reconstruction results for Kanji49 data. Left shows example images from the training set and Right shows example images from the testing data set.*

      For each data set, we trained and tested 3 classification networks using 3 different sources of images: original (ground truth) images, raw sensor images and reconstructed images. The training-testing split is the same as before (9:1). For every network, we used the Adam optimizer to train up to 30 epochs. Figure 6 summarizes classification accuracy for all 3 data sets. A schematic to explain the concept of classification directly from the raw sensor and the second method of classification after reconstruction from the raw sensor is also included in Fig. 6(a).

      For original images, all 3 data sets have a very high training and testing accuracy. But on raw sensor images, the network performs worse. Similar to reconstruction results, MNIST (6 classes) has the highest training and testing accuracy in all 3 settings due to its simplicity and smallest number of classes. The blue bars in Fig. 6 indicate that the testing classification accuracy for MNIST is similar whether one uses reconstructed images or raw-sensor data. Figure 7 shows the confusion matrix of classification for the MNIST dataset. In EMNIST (orange bars in Fig. 6), by first reconstructing the images from raw sensor, then classifying, we could improve classification accuracy. However, in KANJI49 dataset (gray bars in Fig. 6), testing accuracy of reconstructed images is lower than that with raw images. We believe that further parameter tuning of the classifier networks will improve all the accuracies.

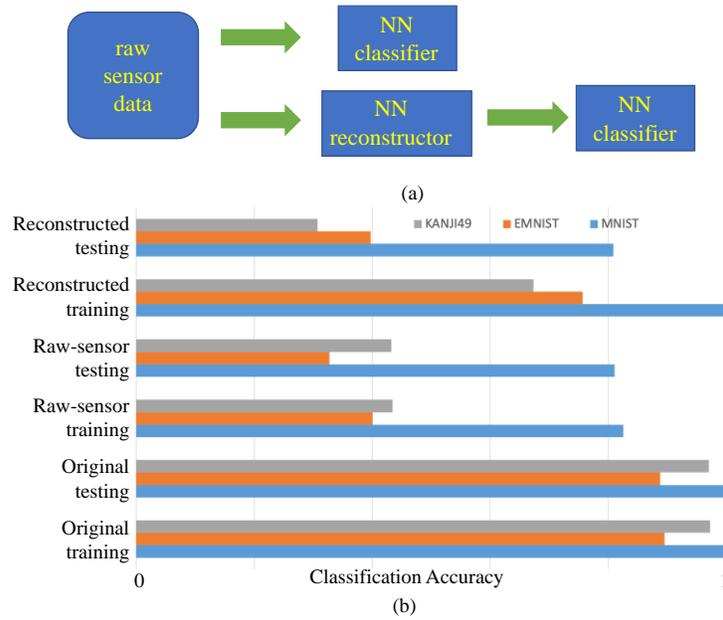

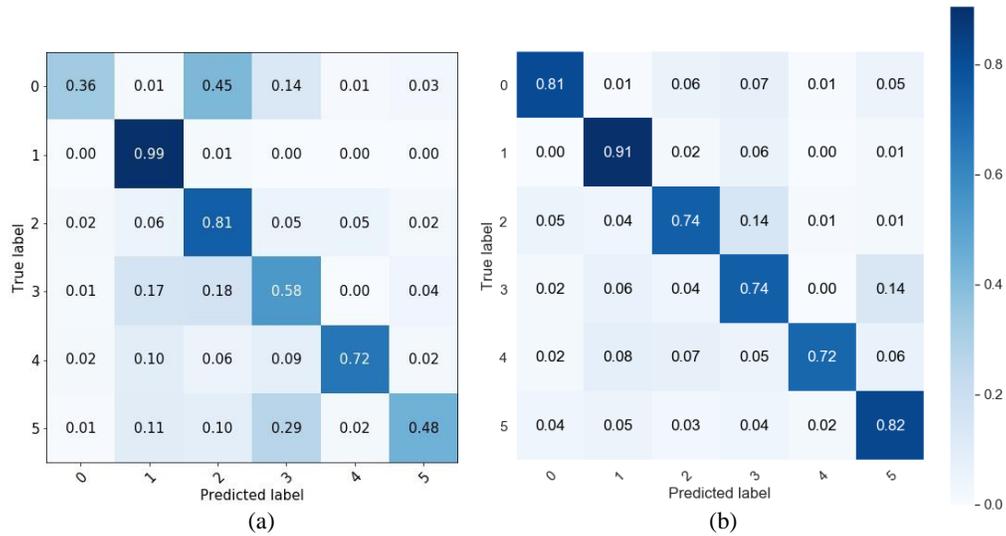

*Figure 6: (a) Schematic of the two methods of classification. (b) Classification accuracy for the two methods and the 3 data-sets.*

*Figure 7: Confusion matrix of classification using (a) raw-sensor images and (b) the reconstructed images from the MNIST (6 classes) dataset.*

In conclusion, we showed that a U-net-based convolutional neural network can be trained to reconstruct images from a "see-through" lensless camera with good fidelity for the MNIST dataset. The quality of the reconstructed images is worse in the case of more complex images as in the EMNIST and the Kanji49 datasets. However, it may be possible to improve these with optimized networks and more training data. Secondly, we compared the accuracy of classification using a standard classifier network using the raw sensor images and the images reconstructed using the U-net. Our conclusion from this preliminary comparison is that for the MNIST dataset with

6 classes, good classification accuracy may be obtained in both cases. However, for more complex data sets like EMNIST and KANJI49, although good image reconstruction is possible, classification accuracy needs further improvement possibly from better network training. We attribute this to the increased complexity of the images as well as the larger number of classes. It must be noted that these results could be improved by optimizing the classifier network architecture for each case separately.

*Funding*
National Science Foundation (NSF) (1533611).

*Acknowledgments*
We would like to thank G. Kim, R. Palmer, A. Kachel and R. Guo for fruitful discussion, and assistance with experiments and software.

*Disclosures*
The authors declare no conflicts of interest.